\documentclass[10pt,conference]{IEEEtran}
\usepackage{balance}

\usepackage{amsthm}
\usepackage{graphicx}
\usepackage{graphics}
\usepackage{multirow}
\usepackage{epsfig}
\usepackage{algorithmic}
\usepackage{algorithm}
\usepackage{subfigure}
\usepackage{amssymb}
\usepackage{amsfonts}
\usepackage{amsmath}
\usepackage{setspace}
\usepackage[normalem]{ulem}
\usepackage{listings}
\usepackage[usenames]{color}

\usepackage{soul}

\usepackage[table]{xcolor}

\begin{document}

%\title{Architecture Design for Massive MIMO-enabled Time-Division Duplex Heterogeneous Networks}
%\title{Flexible TDD Design for Massive MIMO-enabled (Dense) Heterogeneous Networks}
\title{Interference-Aware Flexible TDD Design for Massive MIMO 5G Systems}

\author{
	\IEEEauthorblockN{David M. Gutierrez-Estevez}
	\IEEEauthorblockA{Samsung Electronics R\&D Institute UK, Staines, Middlesex TW18 4QE, UK\\
Email: d.estevez@samsung.com}}
	
%\thanks{Copyright (c) 2015 IEEE. Personal use of this material is permitted. However, permission to use this material for any other purposes must be obtained from the IEEE by sending a request to pubs-permissions@ieee.org}
	
%\thanks{Manuscript received February 23, 2013; revised June 23, 2013, and October 8, 2013; accepted January 11, 2014.}

\maketitle

\begin{abstract}
Both the use of very large arrays of antennas and flexible time division duplexing (TDD) designs have become prominent features of next generation 5G cellular systems. However, both enabling technologies suffer from severe interference effects, respectively known as pilot contamination and base-station-to-base-station (B2B) interference. In this paper, a practical novel TDD design principle is proposed for massive multiple-input multiple-output (MIMO) heterogeneous networks (HetNets) that leverages the inherent features of a flexible TDD design to mitigate both the beamformed interference caused by the pilot contamination effect and B2B interference. The design is based on the key observation that the transmission path chosen for training by the non-massive MIMO base stations plays an important role in the interference behavior of the network, and the data slots need to be configured accordingly. We propose TDFLEX, a low-complexity heuristic solution that follows these design guidelines. Performance evaluation results show significant gains when our design is compared to the standard TD-LTE.
\end{abstract}

%\begin{IEEEkeywords}
%5G Cellular Systems, Massive MIMO, Flexible TDD, Heterogeneous Networks, Pilot Contamination, Interference Regimes, TDD HetNet Grid.
%\end{IEEEkeywords}

%\IEEEpeerreviewmaketitle

\section{Introduction}
\label{sec:intro}

The challenge of ever-increasing demands for wireless data is shaping the design, standardization, and development of systems beyond 4G such as Long Term Evolution-Advanced Pro (LTE-Advanced Pro) and 5G systems \cite{LTEASurvey}, \cite{5G_s2}. For these advanced systems, the use of very large arrays at base stations has been proposed as a very promising way of drastically improving the network capacity \cite{MassiveMIMO1}. However, massive MIMO cellular systems suffer from a fundamental interference limitation known as pilot contamination that has its origin in the necessary reuse of non-orthogonal pilot signals across cells. A common approach is to rely on precoding to mitigate it: A single-cell precoding method is proposed in \cite{PC1}, and multi-cell cooperation methods can be found in e.g. \cite{PC2}. In general, all currently existing approaches must invest resources to combat the problem, usually a non-negligible number of antennas. Finally, the piece of work closest to our approach is the proposal of time-shifted pilots as a means to reduce pilot contamination \cite{PC7}, but only a scenario consisting of macrocells is investigated. 

The applicability of frequency-division duplexing (FDD) in massive MIMO systems has been shown to be challenging because of the amount of pilot overhead and feedback that would be needed for channel estimation \cite{PC1}. Although an interesting debate is taking place in the community on the feasibility of FDD for massive MIMO systems, the feature of channel reciprocity makes time-division duplexing (TDD) very appealing for massive MIMO systems \cite{hoydis2013making}. Furthermore, flexible TDD designs are being proposed as a means to modify the capacity split between uplink and downlink and increase spectrum flexibility, but such schemes introduce the problem of strong base-station-to-base-station (B2B) interference when a downlink transmission happens at the same time of an uplink transmission: The strong transmit power of a macrocell base station (MBS) and a line of sight between base stations may cause unbearably degrading interference \cite{FlexTDD}. 

To the best of our knowledge, proposed flexible TDD designs have not considered massive MIMO deployments where pilot contamination effect is present. In this paper, we tackle this particular challenge by leveraging the characteristics of a flexible TDD architecture to avoid the reception of signal when beamformed interference is directed towards that receiver while mitigating B2B interference. More concretely, we propose design rules based on adjusting the transmission path over which the small cell tier performs training (i.e., downlink or uplink) and configuring the data time slots accordingly so that the interference is minimized. Moreover, we propose TDFLEX, a low-complexity heuristic implementing the above design rules.

The remainder of this paper is organized as follows. In Section \ref{sec:model} we present the system model for this paper. Section \ref{sec:interference} introduces our proposed flexible TDD design principles and identify the different interference regimes. Section \ref{sec:grid} presents the TDFLEX heuristic, and performance evaluation is carried out in Section \ref{sec:perf_eval}. Finally, conclusions are drawn in Section \ref{sec:conc}.

%On the standardisation side, the Third Generation Partnership Project (3GPP) has been making progress on the TDD version of LTE  (TD-LTE) to be applied in future LTE-Advanced and 5G systems. The configuration was initially static and equal for all cells of the network. Currently, a proposed feature of LTE Rel-13 to achieve higher spectrum flexibility is enhanced Interference Mitigation and Traffic Adaptation (eIMTA), which allows for very dynamic adaptation of the TDD pattern according to varying capacity requirements in uplink and downlink \cite{4GAmericasLTEA}.

\section{System Model}
\label{sec:model}
It is assumed that all cells utilize Orthogonal Frequency-Division Multiplexing (OFDM). The channel model for each subcarrier is flat-fading and independent across users and base stations. The estimated channel vector from network element $i$ to $j$ is denoted as $\widehat{\mathbf{h}}_{i}^j$. If the channel is MIMO, we represent it with a matrix $\mathbf{H}$. The pathloss effect is modeled using the standard pathloss function given by $\alpha_i^j = \Vert d(i,j) \Vert^{-\alpha_e}$, where $d(i,j)$ is the distance between network elements $i$ and $j$, and $\alpha_e > 2$ corresponds to the pathloss exponent. We further assume a block fading model both in time and frequency. Channel reciprocity is assumed for both multipath fading and pathloss. Pilot-based channel estimation is assumed in this work. Using channel reciprocity and the coherence block assumption, we safely assume that pilots will be sent in each cell once every $N$ time slots and just over one transmission path. Pilots can only be sent in the uplink for massive MIMO MBSs; SBSs have freedom to choose the transmission path for training. The number of time slots dedicated to uplink ($n_U$) and downlink ($n_D$) communication will depend on the load distribution of the cell, which is assumed to be dynamic.

We assume that the same set of $K$ orthogonal pilots of length $K$ exists for each cell regardless of its tier. No coordinating mechanism among base stations is assumed, thus pilots from different base stations at each tier may be subject to contaminate each other. Each of the $k^{th}$ user in every cell uses the same pilot sequence, namely $\mathbf{\psi}_k = (\psi_{k1}, \psi_{k2}, \cdots , \psi_{kK})$ such that $\vert \psi_{k} \vert = 1$. Since the pilots are orthogonal, we have that $\vert \mathbf{\psi}'_k \mathbf{\psi}_k' \vert = \delta_{k,k'}$.

The multiple antennas of the base stations will be used to perform beamforming in the downlink. The precoding vector is obtained by calculating the normalized version of the estimated channel and the multiple antennas are exploited at the receiver of the base station by decoding the signal using Maximal Ratio Combining (MRC), as performed in \cite{PC7}. Furthermore, we denote by $K_b$ the number of users associated to base station $b$. For data transmission, we denote by $q_k$ the symbol sent by user $k$ in the uplink, and by $s_k^b$ the symbol sent by base station $b$ to user $k$ in the downlink.

The two-tier HetNet considered in this work is comprised of a macrocell and small cells, where the MBS is equipped with a very large number of antennas, the small-cell base stations (SBS) have regular MIMO transmitters, and the users are single-antenna. The total number of SBS in the network is known as $L$. The two tiers are assumed to be deployed using a universal frequency reuse approach. SBSs are distributed following the usual Poisson Point Processes (PPPs). We denote the SBS density as $\lambda_S$. Similarly, the users are also distributed following one single PPP distribution across the network. We denote its density by $\lambda_U$.  
%\begin{figure}[htbp]
     %\centering
     %\includegraphics[width=81mm,height=45mm]{./Figures/Revised-figs/HetNet_v2.pdf}
     %\vspace{-0.9cm}
     %\caption{Considered HetNet topology}
     %\label{fig:hetnet}
%\end{figure}
\section{Novel Flexible TDD Design Proposal and Interference Regimes}
\label{sec:interference}

We are proposing a practical novel design principle for flexible TDD architectures that mitigates the interference caused by the pilot contamination effect in a two-tier HetNet with a massive MIMO macrocell and MIMO small cells. Our contribution leverages the flexibility offered by a dynamic TDD architecture by finding answers to the following two questions:

\begin{itemize}
\item Which transmission path (U or D) should be used for training at the small cell tier?
\item In which order should U/D slots be allocated to prevent pilot contamination while matching the load distribution?
\end{itemize}

In addition, our solution incorporates power considerations that need to be taken into account in any flexible TDD design as a result of the B2B interference problem.

\subsection{Basic Idea}
\label{ssec:idea}

Each frame consists of $N+1$ time slots ($N$ for data, $1$ for training), each of which must be configured in one of the following modes of operation: 
\begin{itemize}
\item Uplink Pilot ($S_U$), for reference signals transmitted in the uplink.
\item Downlink Pilot ($S_D$), for reference signals transmitted in the downlink.
\item Uplink ($U$), for uplink data transmission.
\item Downlink ($D$), for downlink data transmission.
\end{itemize}

However, we impose in our design the following constraints while seeking for a suitable configuration of the TDD HetNet: i) Pilots in the macrocell tier must be sent on the uplink ($S_U$) due to massive MIMO overhead constraints; ii) pilots in the small cell tier may be sent both on the uplink ($S_U$) or downlink ($S_D$); iii) the number of slots for data transmission in U/D should be computed based on the load distribution while the slot location within the frame is freely selected; and iv) the selection of $S_U/S_D$ for the cells in the small cell tier and the location of the data slots ($U/D$) within the frame will be performed with the objective of minimizing the impact of the pilot contamination effect.

\begin{figure}[htb]
     \centering
     \includegraphics[width=0.50\textwidth]{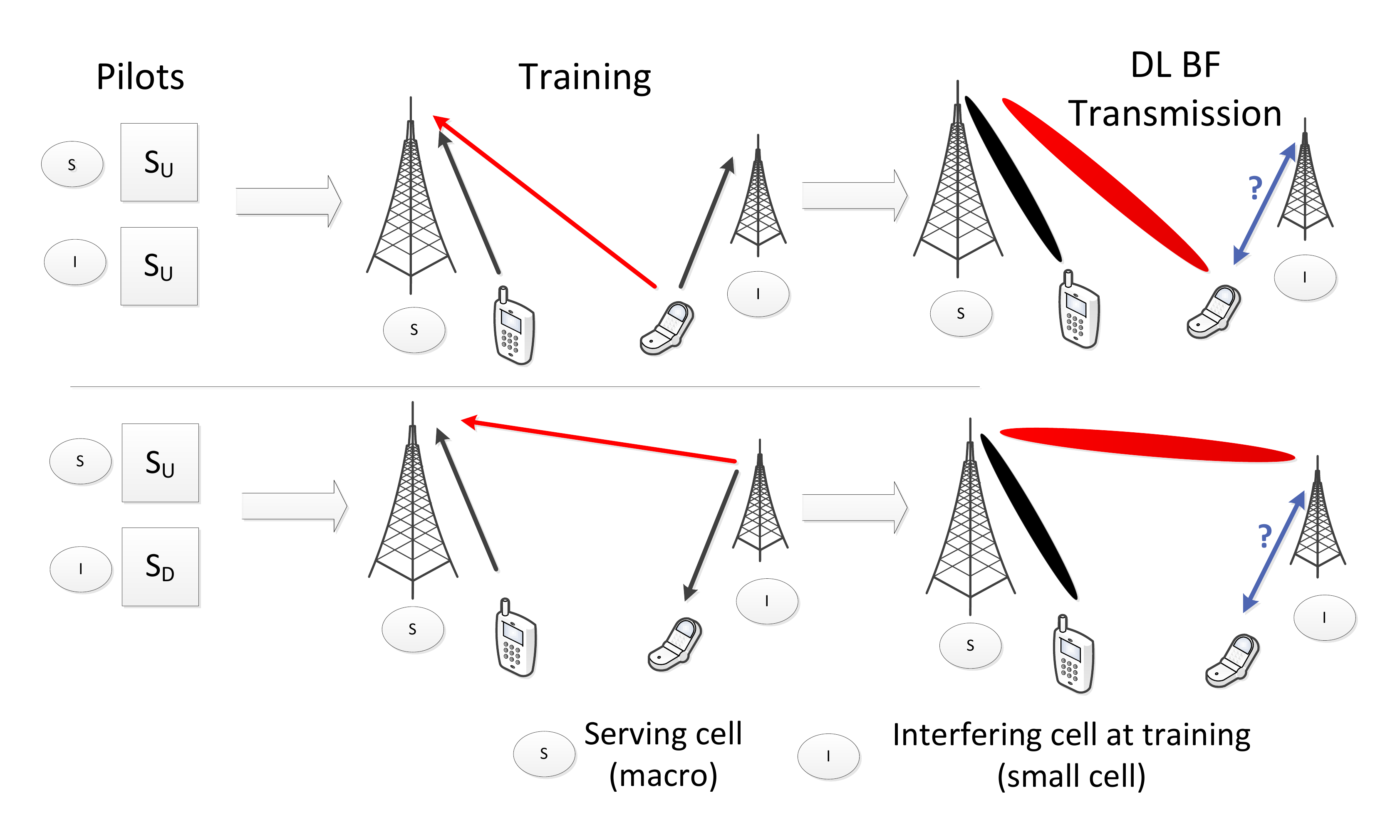}
     \vspace{-0.9cm}
     \caption{Sketch of the basic idea.}
     \label{fig:basic}
\end{figure}

Given the above principles, the basic idea of our TDD design proposal is summarized in Fig. \ref{fig:basic}. The pilot of the macrocell may be contaminated by $S_D$ or $S_U$, depending on the transmission path that is used for training at the small cell. This creates two possible cases of beamformed interference, either directed to the user or to the SBS. In that context, the small cell needs to decide which transmission path to use during that time slot. The obvious answer is the configuration that avoids receiving signal at a network element while the beamformed interference is being directed to that same network element. However, the freedom to choose is limited by the load distribution, which imposes the number of U and D slots in the cell. How to solve this problem will be shown in the following. 
\subsection{Interference Regimes}
Different interference regimes between the macrocell and each of the small cells may appear as a consequence of the pilot contamination effect. An interference regime is the result of (i) the existence of pilot contamination effect, ii) the transmission path used for training at the small cell, and iii) the transmission path chosen for data transmission at both the macrocell and the small cell. When an interfering beam is generated as a result of pilot contamination effect we consider to be operating under \textbf{Pilot Contamination Regime (PCR)}. We do not consider existence of interference regimes between small cells as we assume the likelihood of their users to use the same pilot signal is much smaller and they are not equipped with a very large number of antennas.

The case of PCR implies that the channel estimate is contaminated in such a way that beamforming the transmit signal at the base station will imply directing interference to another existing element in the network. For that effect to happen, the interference received during the training phase of the macrocell must come from a small cell (in uplink or downlink) that is using the same pilot signal as the particular user being served in the same time slot. As previously explained, the mode $S_U$ is used during the training phase at the macrocell for overhead reasons. This means that the estimation is performed at the base station, and the contaminated estimates correspond to the (reciprocal) channels between the receiving base station and the interfering element. Two possibilities exist for the interfering element: A user or a base station. We call the first regime \textit{PCR-U} and the second \textit{PCR-D}. The former implies that the interfering cell operates in $S_U$ mode while the latter implies an $S_D$ mode during the training phase. 
%Assuming that during PCR-U all interfering cells are operating in $S_U$ mode, the received signal at base station $b$ can be expressed as follows:
%\begin{equation}
%\mathbf{y}_{b}^{PCR-U} = \mathbf{r}_{b}^{S_U} + I_{b}^{S_U} + \mathbf{z},
%\end{equation}
%where $\mathbf{z}$ is the additive white gaussian noise (AWGN) and $I_{b}^{S_U}$ represents one of the four components of $I_x$ in \eqref{eq:Ix} when $x$ stands for a base station. Then, 

In PCR-U, the pilot-based channel estimation for user $k$ at base station $b$ is performed by multiplying the received signal $\mathbf{y}$ with the pilot as follows:
\begin{equation}
\label{eq:PU_IPU}
\begin{split}
\widehat{\mathbf{h}}_{k}^{b} &= \mathbf{\psi}'_{k} \mathbf{y}_{b}^{PCR-U} \\
&= \sqrt{\alpha_{k}^{b}P^u_{k}}\mathbf{h}_{k}^{b} + \sum_{b'=1}^{B^{k}}\sqrt{\alpha_{k'_{b'}}^{b'}P^u_{k'_{b'}}}\mathbf{h}_{k'_{b'}}^{b} + \mathbf{z},
\end{split}
\end{equation}
where $\mathbf{z}$ is AWGN, $P^u_k$ is the transmit power of user $k$, $B^{k}$ is the number of cells using the pilot sequence $\mathbf{\psi}_k$ for channel estimation, $k'_{b'}$ is the particular user in the interfering cell $b'$ using the pilot sequence $\mathbf{\psi}_k$, and the pilot orthogonality property has been used to cancel the the rest of terms. The most interesting element of Eq. \eqref{eq:PU_IPU} is the channel vector $\mathbf{h}_{k'_{b'}}^{b}$, which generically represents \emph{the channels from all interfering users using the same pilot sequence to the serving base station $b$}. These particular channels that have contaminated the estimates become very degrading interference when beamforming is performed since the precoding vector is just the normalized version of the channel estimate.

Let us now look at the PCR-D, where the channel estimate performed for user $k$ at base station $b$ is given by
%case assuming all interfering cells are operating in $S_D$ mode. The received signal at base station $b$ is given by
%\begin{equation}
%\mathbf{y}_{b}^{PCR-D} = \mathbf{r}_{b}^{S_U} + I_{b}^{S_D} + \mathbf{z},
%\end{equation}
%and the channel estimate performed for user $k$ at base station $b$ is given by
\begin{equation}
\begin{split}
\widehat{\mathbf{h}}_{k}^{b} &= \mathbf{\psi}'_{k} \mathbf{y}_{b}^{PCR-D} \\
&= \sqrt{\alpha_{k}^{b}P^u_{k}}\mathbf{h}_{k}^{b} + \sum_{b'=1}^{B^{k}}\sqrt{\alpha_{k'_{b'}}^{b}P_b}\left(\mathbf{\psi}'_{k}\bar{\mathbf{H}}_{i}^{b'}\mathbf{\psi}_{k}\right) + \mathbf{z} \\
&= \sqrt{\alpha_{k}^{b}P^u_{k}}\mathbf{h}_{k}^{b} + \sum_{b'=1}^{B^{k}}\sqrt{\alpha_{k'_{b'}}^{b}P_b}\left(\bar{\mathbf{H}}_l\right)_{b'}^{b} + \mathbf{z},
\end{split}
\end{equation}
where $P_b$ is the transmit power of base station $b$, $\left(\bar{\mathbf{H}}_l\right)_{b'}^{b}$ \textit{represents the $l$-th SIMO channel from the interfering base station $b'$ to the serving base station $b$ using the same pilot signal as user $k$}. Hence, we observe a major difference between PCR-U and PCR-D modes: The contaminating contributions represent a base station to user (B2U) channel in the case of PCR-U and a B2B channel in the case of PCR-D. 

The above observation has important implications in the TDD design. The main directivity gains of beamforming come from pre-multiplying the transmit signal by the same channel vector that will be undergone by the signal before being received, hence increasing its energy in that particular direction. Precoding the signal with a channel estimate that contains the interfering channel during the training phase will boost the transmission in that particular direction, increasing interference. However, the proposed TDD architecture provides the flexibility to avoid the reception of signal when a strong interfering beam is directed towards the receiver by selecting the appropriate transmission paths configuration given the training phase configuration of the cells. 

Let us call \emph{Reduced Contamination Regime (RCR)} the TDD configuration that avoids listening when directed interference exists, and \emph{Increased Contamination Regime (ICR)} the case when beamformed interference is received. The problem thus reduces to finding the RCR and ICR configurations for both PCR-U and PCR-D regimes. In the case of PCR-U, we need to avoid the user who contaminated the pilot to be listening to the channel when the serving base station transmits. Hence, the $D$ mode should be avoided in the interfering cell when $D$ is selected in the serving cell. Similarly, when the serving cell is receiving data in $U$ mode, the $U$ mode should not be utilized in the interfering cell. Hence, these two configurations represent ICR in PCR-U. An analog reflection follows for PCR-D regime, as Section \ref{ssec:rules} further clarifies.

\subsection{Power Considerations}
\label{ssec:power}

As mentioned above, flexible TDD designs suffer from the problem of B2B interference: When two interfering cells are configured with different transmission paths in the same time slot (i.e., in reverse TDD mode), the cell operating in $D$ mode could cause very strong interference to the cell in $U$ mode as base stations have very large transmit powers in comparison to users and they might be located within line of sight. %Several approaches have been proposed in the literature to deal with this problem such as enhancing the uplink powers of the cell operating in $U$ mode or clustering cells in such a way that interfering base stations are configured in the same modes \cite{FlexTDD}. %In the case of a HetNet, within the coverage region of a single macrocell it is more likely that the MBS causes strong interference to small cells than vice versa due to its stronger transmit power. Furthermore, small cells may not be necessarily surrounded by other small cells, and if they are, a simple cell clustering solution among the SBS subject to interference would prevent them from causing B2B interference to each other.
%In this work, we focus on the prevention of beamformed interference caused by the MBS to the small cells due to the pilot contamination effect. 
The pilot contamination degradation becomes thus intolerable if interference is both beamformed and boosted as in the B2B case. Hence, power considerations will also play a role in the classification of the different interference regimes identified in the previous section. In particular, \emph{the case of B2B macro-to-small-cell interference while in ICR regime must be avoided at all costs} as it contains both pilot contamination and B2B interference. Furthermore, B2B macro-to-small-cell interference while in RCR should be tackled even if no beamformed interference is present. For that case, we choose to utilize the existing technique of enhancing the uplink power of small cell users as a means to compensate for the B2B interference coming from the MBS.

\subsection{Proposed Flexible TDD Design Rules}
\label{ssec:rules}

Table \ref{tb:main_results} shows the key findings of this paper, where each table cell represents a transmission path ($D$ or $U$) of the macrocell or a small cell in one single data slot. The table is structured as two sets of two rows each, each set (PCR-D and PCR-U) representing an interference regime resulting from the transmission path that contaminated the pilot estimation: Downlink in case of PCR-D and uplink in case of PCR-U. Furthermore, each row in each set represents a cell's TDD configuration during data transmission slots, where the S rows represent the serving macrocell configuration and I rows represent the interfering small cells. The columns identify which of the TDD configurations correspond to a well-managed pilot contamination case (RCR) that avoids beamformed interference and which correspond to an increased contamination regime (ICR). Moreover, the power considerations presented in Section \ref{ssec:power} are also taken into account as colors to specify our proposed rules for a TDD design.

\begin{table}[h]
\centering
\caption{Pilot Contamination Regime Classification}
\begin{tabular}{cc|c|c|c|c|}
\cline{3-6}
& & \multicolumn{2}{|c|}{RCR} & \multicolumn{2}{|c|}{ICR} \\ \hline
\multicolumn{1}{ |c| }{\multirow{2}{*}{PCR-D} } &
\multicolumn{1}{ |c| }{S} & \cellcolor{yellow!25}D & \cellcolor{yellow!25}U & \cellcolor{red!25}D & \cellcolor{red!25}U     \\ \cline{2-6}
\multicolumn{1}{ |c  }{}                        &
\multicolumn{1}{ |c| }{I} & \cellcolor{yellow!25}D & \cellcolor{yellow!25}U & \cellcolor{red!25}U & \cellcolor{red!25}D     \\ \cline{1-6}
\multicolumn{1}{ |c  }{\multirow{2}{*}{PCR-U} } &
\multicolumn{1}{ |c| }{S} & \cellcolor{green!25}D & \cellcolor{green!25}U & \cellcolor{blue!25}D & \cellcolor{blue!25}U \\ \cline{2-6}
\multicolumn{1}{ |c  }{}                        &
\multicolumn{1}{ |c| }{I} & \cellcolor{green!25}U & \cellcolor{green!25}D & \cellcolor{blue!25}D & \cellcolor{blue!25}U \\ \cline{1-6}
\end{tabular}
\label{tb:main_results}
\end{table}

A division by quadrants of Table \ref{tb:main_results} visually allows a prioritized classification of the different possible TDD modes. RCR is obviously preferred over ICR for the pilot contamination reasons previously stated. Within RCR, the PCR-D mode (yellow) does not require any modification of the transmit powers while PCR-U (green) requires to enhance the uplink transmit power of small cell users to counteract the B2B reverse TDD interference. In the case of ICR, the PCR-U mode (blue) suffers from beamformed interference but it is preferred over PCR-D (red) as the latter adds B2B interference to pilot contamination. Hence, the TDD configuration of the cells in a HetNet must be set following this colored priority order: \emph{1) yellow, 2) green, 3) blue, and 4) red}. This categorization provides powerful insights on how a TDD HetNet architecture should be designed to mitigate the critical pilot contamination effect of massive MIMO systems.

\begin{figure}[htb]
     \centering
     \includegraphics[width=0.35\textwidth]{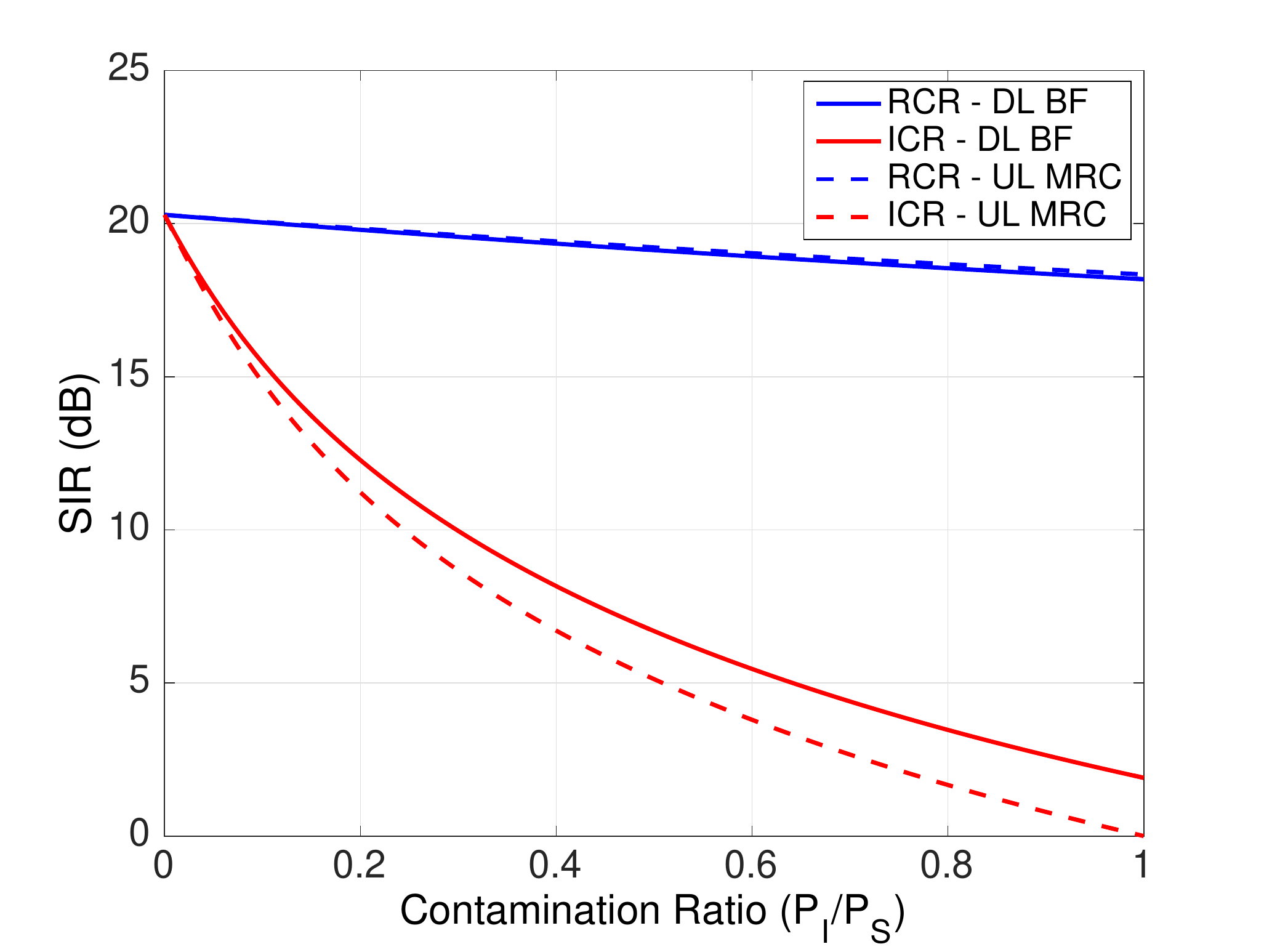}
     %\vspace{-0.5cm}
     \caption{Pilot Contamination Regime analysis.}
     \label{fig:PCR_new}
\end{figure}

We now present a PCR analysis where performance of RCR and ICR are compared. We evaluate the simple case of two cells with one user each sharing the pilot signal where one cell acts as the serving cell and the second cell as interferer. The serving base station is equipped with a very large array of antennas. We restrict the simulation to two time slots: A training phase followed by a data transmission phase. We further assume that the serving base station gets its channel estimate contaminated by interfering pilots carried in the downlink (PCR-D). Then, we measure the signal-to-interference ratio (SIR) during the data transmission slot at the receiving ends, namely the interfered user in the case of downlink data transmission, and the serving base station in the case of uplink transmission. The results are displayed in Fig. \ref{fig:PCR_new}. The SIRs are measured for different contamination ratios, defined as the quotient between the received serving power and the received interfering power during the training phase. Clearly, the pilot contamination effect degrades SIRs both in the downlink and uplink. More interestingly, selecting the RCR configuration over ICR greatly increases the SIR of the downlink and uplink transmissions. Furthermore, the contamination ratio plays an important role: When the power level of the contamination is high, the beamformed interference experienced at the users increase, hence enlarging the SIR gap between RCR and ICR. This observation is crucial when designing a TDD configuration for a HetNet since beamformed interference coming from high-power elements is much more dangerous than the interference coming from low-power elements. In summary, we have observed that the design of the TDD configuration is a critical parameter to control interference in massive MIMO systems.

\section{The TDFLEX Heuristic}
\label{sec:grid}
In this section, we we propose a heuristic solution to the problem of configuring the TDD frames of a HetNet called TDFLEX. The problem can be described as follows: Given a separate dynamic load for each cell, we need to dynamically fill out the time slots of the TDD frames with the objective of minimizing the interference caused by the pilot contamination effect. Assuming training occurs in the first time slot of every frame, a main constraint is the need for an $S_U$ mode during the training slot of the massive MIMO MBS. Then, the rest of training slots at every small cell must be set to either $S_U$ or $S_D$. Furthermore, data slots need to be filled with either $U$ or $D$ modes to minimize pilot contamination (i.e., induce the best possible configurations according to Table \ref{tb:main_results}). 
\begin{table}[!htb]
\centering
\caption{TDFLEX Parameters}
\begin{tabular}[c]{|c|c|}
\hline
\textbf{Element} & \textbf{Description} \\ \hline
$N$ & Data time slots in a frame \\ \hline
$L$ & Small cells \\ \hline
$n_D$ & DL time slots \\ \hline
$n_U$ & UL time slots \\ \hline
$l_D, l_U$ & Load distribution percentage \\ \hline
$\gamma$ & UL power enhacement parameter \\ \hline
$\{\beta_{SU}, \beta_{SD}, \beta_U, \beta_D \}$ & $\{S_U, S_D, U, D \}$ \\ \hline
\end{tabular}\label{tb:tdd_grid}
\end{table}
The TDFLEX is a low-complexity heuristic that dynamically matches the load distribution while minimizing the number of ICR collisions $C$. For that, it induces either PCR-D or PCR-U modes, whichever is more suitable, with the purpose of minimizing the impact of the pilot contamination effect. The working principle of TDFLEX is as follows: By sorting the $U$ and $D$ slots, the number of collisions both in PCR-D and PCR-U modes can be easily computed given the load. In particular, \eqref{eq:c_pcrd} provides the number of collisions for a given load distribution if PCR-D is chosen, while \eqref{eq:c_pcru} does the same in the case of PCR-U. In both equations, the superscript M represents the macrocell and S the contaminating small cell. In addition, power considerations are taken into account to favor PCR-D when no collision exists while favoring PCR-U if a collision is unavoidable. Once the computations are performed and the slots filled, columns of data slots in the $\mathbf{A}$ matrix can be rearranged if certain slot sorting is preferred. The complexity of the algorithm is $\mathcal{O}(LN)$, i.e., linear in the product of number of cells times number of time slots within a subframe. This means that a huge complexity reduction is achieved in comparison with the exact optimization problem.
\begin{equation}
\label{eq:c_pcrd}
C^{PCR-D} = N - \left[\min \left(n_D^M, n_D^S\right) + \min \left(n_U^M, n_U^S\right) \right]
\end{equation}
\begin{equation}
\label{eq:c_pcru}
C^{PCR-U} = N - \left[\min \left(n_D^M, n_U^S\right) + \min \left(n_U^M, n_D^S\right) \right]
\end{equation}
The complete TDFLEX algorithm is shown below. The main steps of TDFLEX are summarized as follows:
\begin{itemize}
\item Calculate $n_U$ and $n_D$ for each cell.
\item Fill out the pre-sorted macrocell TDD row of $\mathbf{A}$.
\item For each small cell:
\begin{itemize}
\item Calculate collisions with PCR-D using \eqref{eq:c_pcrd}. If B2B interference appears, discard.
\item Calculate collisions with PCR-U using \eqref{eq:c_pcru}.
\item Select mode with lesser number of collisions, prioritizing PCR-D for equal Cs as less power is needed.
\item Fill out sorted data slots accordingly
\item If PCR-U is selected, enhance uplink power in U slots under RCR.
\item Rearrange data slot columns if desired.
\end{itemize}
\end{itemize}
\begin{figure}[htbp]
\begin{center}
\line(1,0){250}
\end{center}
\vspace{-0.3cm}
%\hspace{0.1cm} 
\begin{center}
\textbf{TDFLEX Heuristic}
\end{center}
\vspace{-0.6cm}
\begin{center}
\line(1,0){250}
\end{center}
\begin{center}
\begin{algorithmic}[1]

\STATE INPUT: $N$, $\mathbf{l_D}$, $\mathbf{l_U}$, $\gamma$ 
\STATE OUTPUT: $\mathbf{A}$
\STATE DEFINE: $S_U=\beta_0$, $S_D=\beta_1$, $U=\beta_2$, $D=\beta_3$ 
\FOR {$b \leftarrow 0, B-1$}
\STATE $n_D[b] \leftarrow $round$\left( \frac{(N-1)l_D[b]}{l_D[b]+l_U[b]}\right)$
\STATE $n_U[b] \leftarrow N-1-n_D[b]$
\ENDFOR

\STATE $A[0][0] \leftarrow S_U$
\STATE $A[0][1:n_D[0]] \leftarrow D; A[0][1:n_U[0]] \leftarrow U$
\FOR {$b \leftarrow 1, B-1$}
\STATE $C^{PCR-D} \leftarrow N - \left[\min \left({n_D}[0], n_D[b]\right) + \min \left(n_U[0], n_U[b]\right)\right]$
\IF{$\left(n_D[0] > n_D[b]\right)$}
\STATE $C^{PCR-D} \leftarrow \infty$
\ENDIF
\STATE $C^{PCR-U} \leftarrow N - \left[\min \left(n_D[0], n_U[b]\right) + \min \left(n_U[0], n_D[b]\right)\right]$
\IF{$C^{PCR-D} \leq C^{PCR-U}$}
\STATE $A[b][0] \leftarrow S_D$ 
\STATE $A[b][1:n_D] \leftarrow D$
\STATE $A[b][n_D+1:N] \leftarrow U$
\ELSE
\STATE $A[b][0] \leftarrow S_D$ 
\STATE $A[b][1:n_D] \leftarrow D$
\STATE $A[b][n_D+1:N] \leftarrow U$
\STATE $P_U^S \leftarrow \gamma P_U^S$
\ENDIF
\ENDFOR

\end{algorithmic}
\end{center}
\vspace{-0.6cm}
\begin{center}
\line(1,0){250}
\end{center}
\label{fig:tdflex}
\end{figure}

\section{Performance Evaluation}
\label{sec:perf_eval}
System-level simulations are conducted on a two-tier HetNet deployed across the area of one single macrocell equipped with a very large antenna array. The rest of the base stations in the network are equipped with two-antenna transmitters and users are single-antenna. Table \ref{tb:sim_parameters} shows the most relevant parameters to simulate the network described in Section \ref{sec:model}. The first system-level evaluation will be performed with an array at the MBS of $M=128$ antenna elements, while the second one will investigate the impact of the number of antennas on the network performance. The objective of this performance evaluation is to obtain the rate distribution of the users potentially affected by the pilot contamination effect using two different design approaches for the TDD configuration: The proposed TDFLEX heuristic and TD-LTE. The uplink power enhancement factor $\gamma$ in TDFLEX is chosen heuristically to maximize performance. The load changes dynamically in each time frame and follows a uniform random distribution. Furthermore, the most suitable TD-LTE reference configuration is chosen for each time frame to match the load requirements, in a similar fashion to what 3GPP's eIMTA enables \cite{NomorFlexTDD}. It is important to point out that this is not a load modeling work, hence the load model is quite simplified and both TDFLEX and TD-LTE are considered to exploit the TDD flexibility to allocate the needed number of uplink and downlink slots. What these experiments will show, however, is the difference in data rates between the two approaches due to interference caused by pilot contamination and unmanaged B2B interference.
\begin{table}[!htb]
\centering
\caption{Relevant Simulation Parameters}
\begin{tabular}[c]{|c|c|}
\hline
\textbf{Parameter} & \textbf{Value}\\ \hline
Macrocell area & 1 $km^2$ \\ \hline
Density of small cells $(\lambda_{SC})$ & $0.6$ \\ \hline
Mobile user density $(\lambda_U)$ & 50 \\ \hline
TX power of MBS, SBS (dBm) & 43, 25  \\ \hline
%TX power of MUEs, PUEs, FUEs (dBm) & (0, -3, -6)  \\ \hline
Pathloss exponent & 3 \\ \hline
%Channel exponential mean & 1 \\ \hline
$N$ & $8$  \\ \hline
$n_D, n_U (\%)$ & $U(0.5), 1-U(0.5)$  \\ \hline
\end{tabular}\label{tb:sim_parameters}
\end{table}

\begin{figure}[h]
     \centering
     \subfigure[Downlink]{\includegraphics[width=0.235\textwidth]{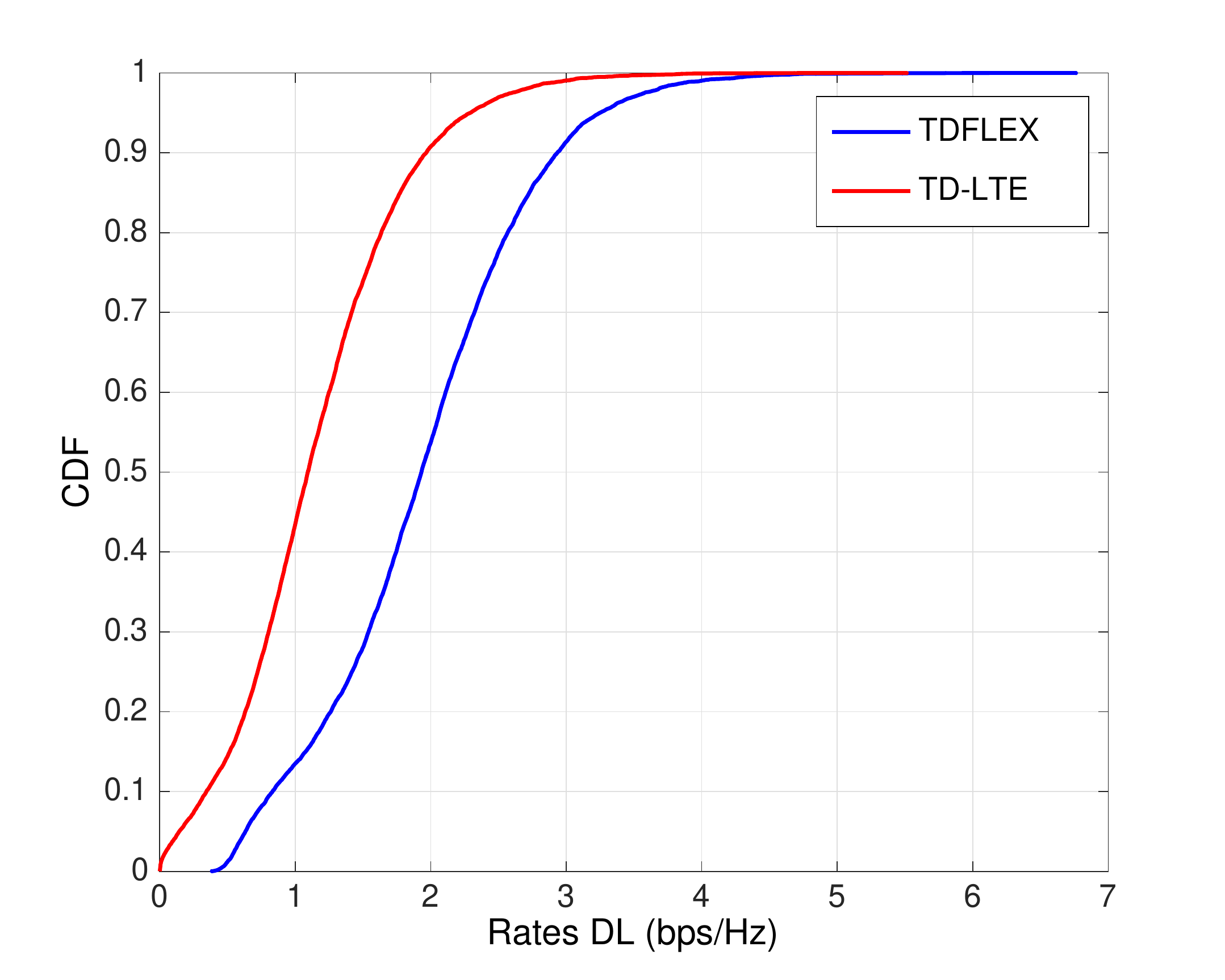}}
     \subfigure[Uplink]{\includegraphics[width=0.235\textwidth]{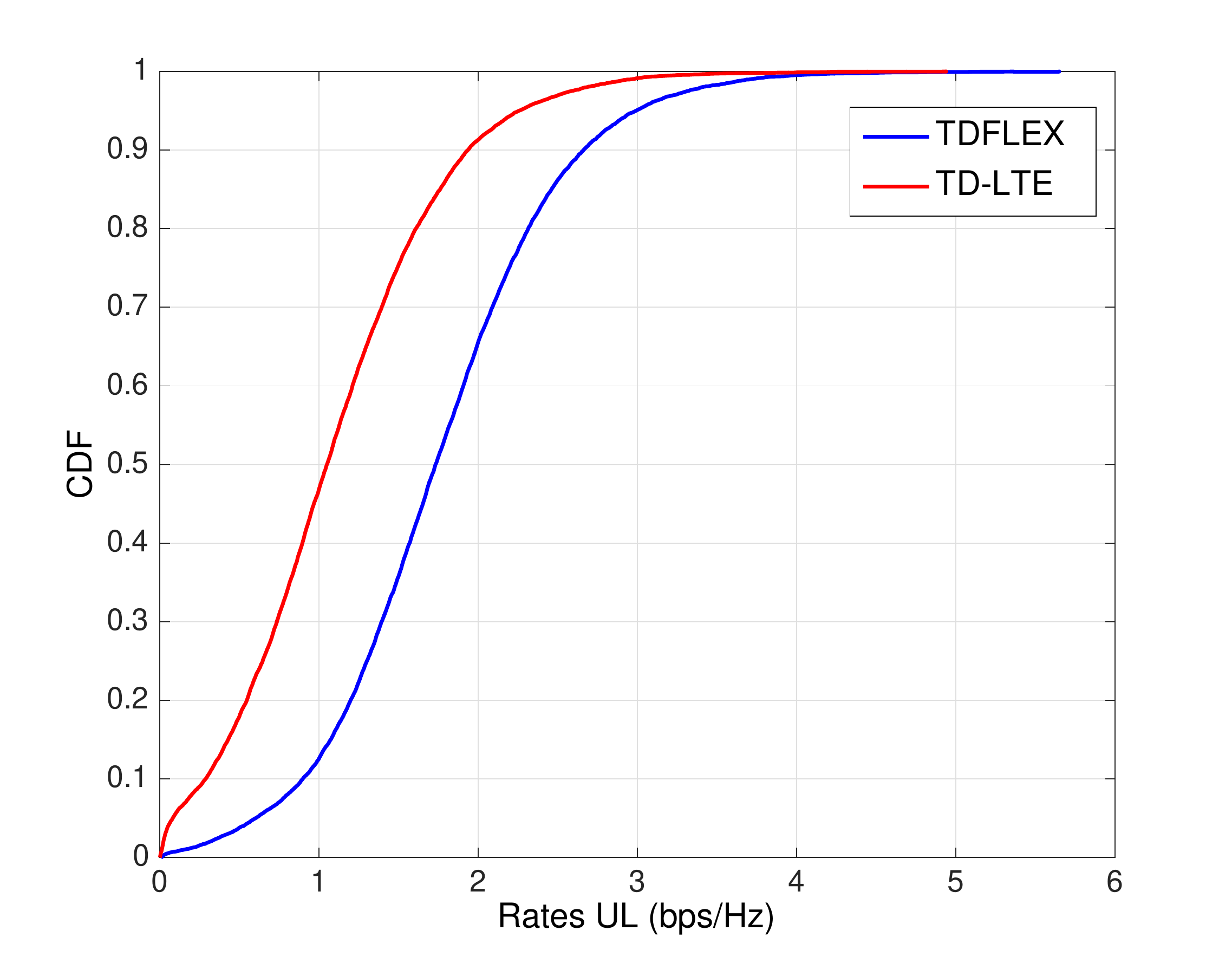}}
		 \caption{Rate distribution for different TDD grid solutions}
     \label{fig:RatesComparison}
\end{figure}

The resulting uplink and downlink rate distributions are shown in Fig. \ref{fig:RatesComparison}, where the rates are expressed as spectral efficiency quantities measured in bps/Hz. Downlink rates represent the rates of the interfered users (i.e., all small cell users) while uplink rates are measured at the MBS since no pilot contamination effect appears in single-antenna base stations. Several interesting observations can be made. First, managing the interference caused by the pilot contamination effect in a HetNet by means of the TDD architecture makes a very positive impact on the attainable user rates: By selecting the appropriate receiver when beamformed interference is present, the user rates can fully benefit from the advantages of massive MIMO systems. This advantage can be easily missed if the TDD architecture is not designed with this objective in mind, as in the case of dynamic TD-LTE. The effect, although troublesome for both downlink and uplink communications, is particularly bad for small cell users who see their rates very limited by the beamformed interference coming from the MBS. 

\begin{figure}[ht]
     \centering
     \subfigure[TDFLEX in downlink]{\includegraphics[width=0.235\textwidth]{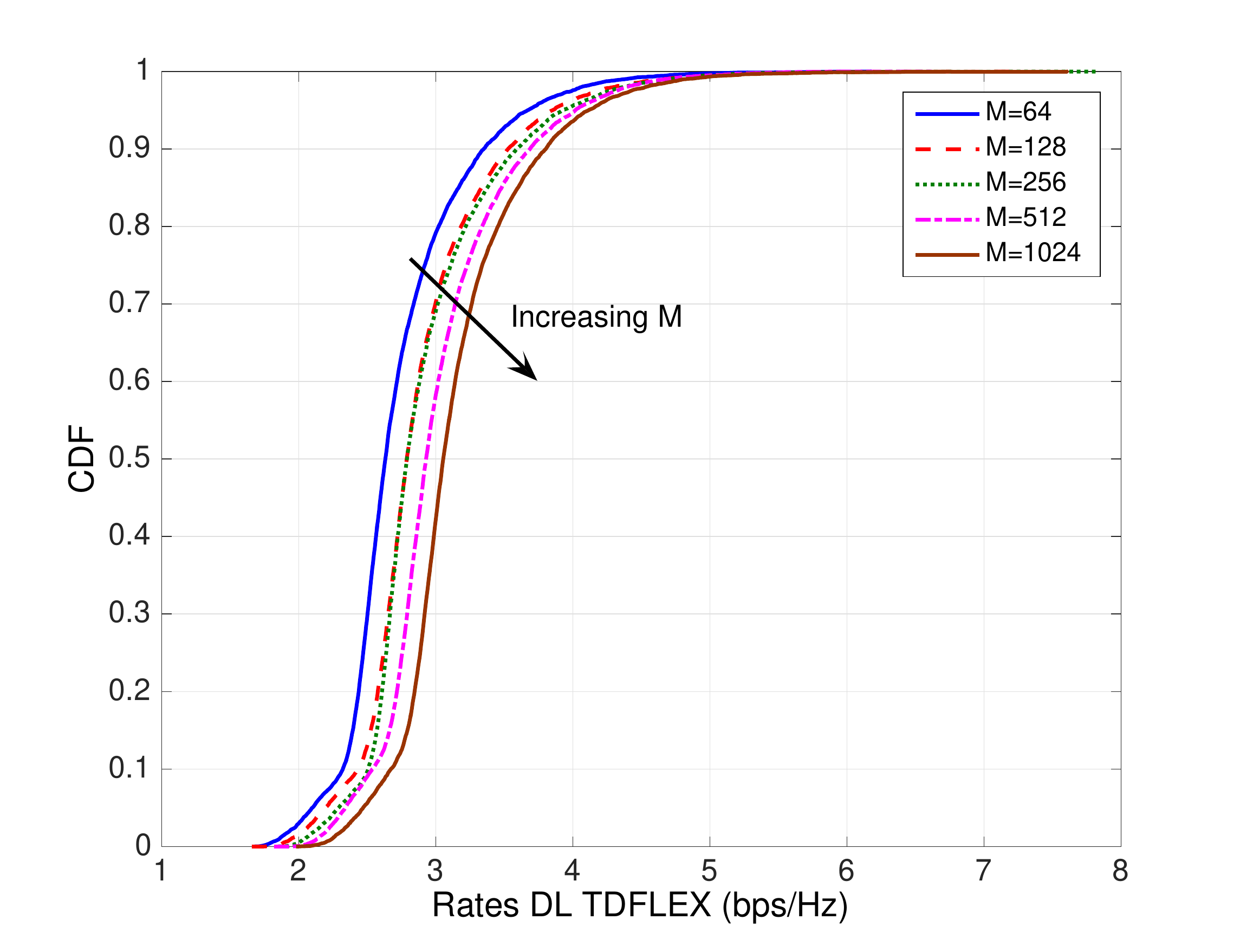}}
     %\subfigure[TDFLEX in uplink]{\includegraphics[width=0.40\textwidth]{./Figures/M_aPCR_UL.pdf}}
     \subfigure[TD-LTE in downlink]{\includegraphics[width=0.235\textwidth]{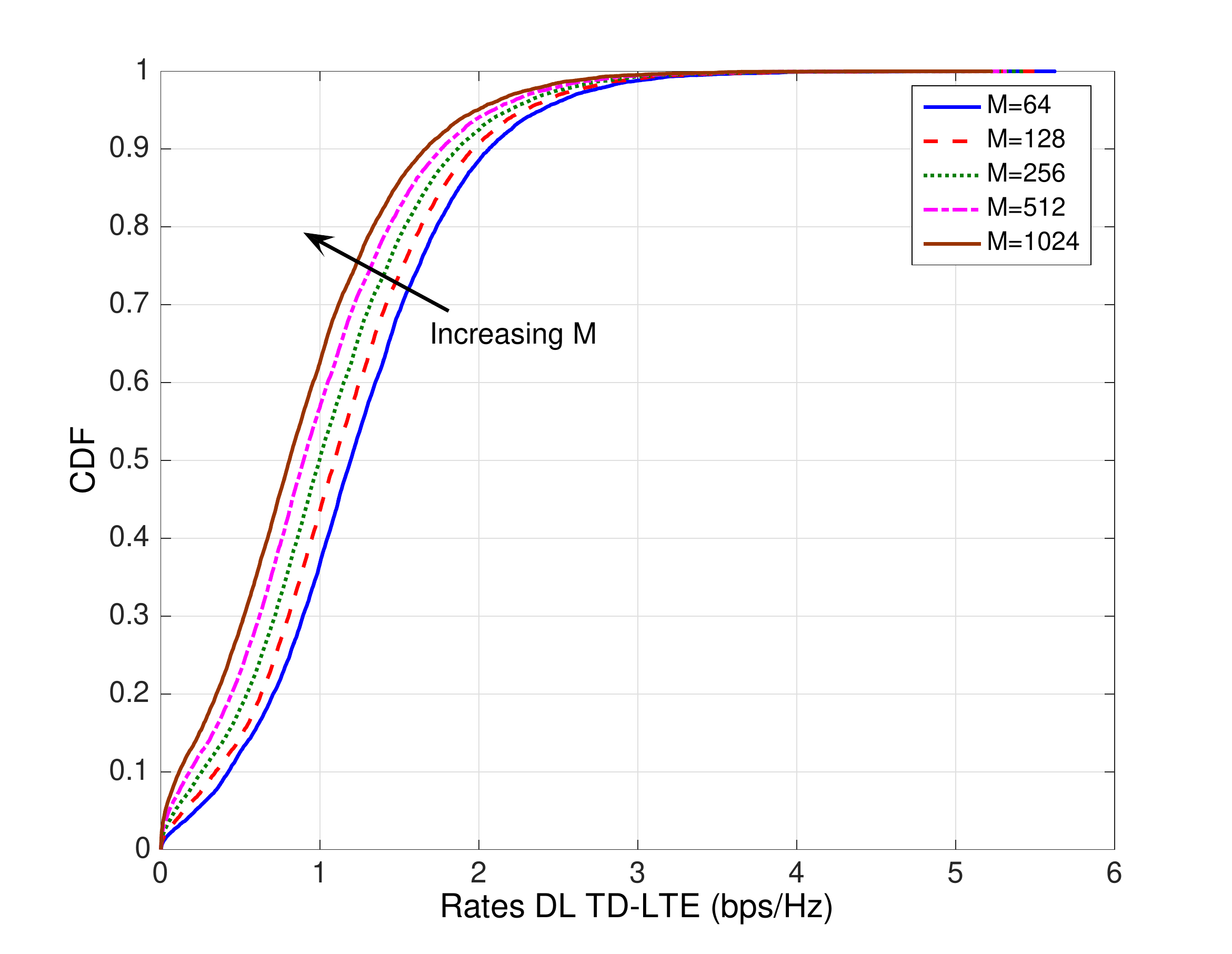}}
     %\subfigure[TD-LTE in uplink]{\includegraphics[width=0.40\textwidth]{./Figures/Revised-figs/M_TDLTE_UL.pdf}}
     \caption{Rate distributions for different number of antennas at the MBS}
     \label{fig:PCR_analysis}
\end{figure}

Finally, we show a very interesting effect observed in the downlink of our massive MIMO-enabled HetNet: A very different performance dependence on the number of antennas for the two presented TDD designs, namely TDFLEX and TD-LTE. The results are shown in Fig. \ref{fig:PCR_analysis}. The most important observation that can be extracted from these results is the difference in the evolution of the rate distribution curve when the number of antennas at the MBS $M$ is increased and beamforming is present. If the pilot contamination effect is well managed via TDFLEX, the users rates benefit from the increase of the number of antennas. However, if the underlying TDD architecture does not account for this effect such as the standard TD-LTE, the increase of antennas can be counter-productive and greatly damage the attainable rates of users subject to receive beamformed interference in the downlink. 

\section{Conclusions}
\label{sec:conc}

In this paper, we present a novel method to mitigate the interference caused by the pilot contamination effect of a massive MIMO HetNet by leveraging a flexible TDD design. Our method does not require to sacrifice antennas to avoid pilot contamination, and it effectively manages B2B interference. The main idea is to prevent the reception of signal at time slots where a network element is receiving strong interference from a transmitter that uses a contaminated channel estimate by wisely selecting the configuration of pilot and data time slots. Results show a significance improvement in performance when the pilot contamination effect is managed via our proposed heuristic TDFLEX against 3GPP's dynamic TD-LTE. Furthermore, our results also show that further increasing the number of antennas brings additional gains only if beamformed interference is avoided.

\section*{Acknowledgment}
Part of this work has been performed in the framework of the H2020 project METIS-II co-funded by the EU. 

\bibliographystyle{IEEEtran}
\bibliography{tdd_ref}

% Generated by IEEEtran.bst, version: 1.14 (2015/08/26)
\begin{thebibliography}{1}
\providecommand{\url}[1]{#1}
\csname url@samestyle\endcsname
\providecommand{\newblock}{\relax}
\providecommand{\bibinfo}[2]{#2}
\providecommand{\BIBentrySTDinterwordspacing}{\spaceskip=0pt\relax}
\providecommand{\BIBentryALTinterwordstretchfactor}{4}
\providecommand{\BIBentryALTinterwordspacing}{\spaceskip=\fontdimen2\font plus
\BIBentryALTinterwordstretchfactor\fontdimen3\font minus
  \fontdimen4\font\relax}
\providecommand{\BIBforeignlanguage}[2]{{%
\expandafter\ifx\csname l@#1\endcsname\relax
\typeout{** WARNING: IEEEtran.bst: No hyphenation pattern has been}%
\typeout{** loaded for the language `#1'. Using the pattern for}%
\typeout{** the default language instead.}%
\else
\language=\csname l@#1\endcsname
\fi
#2}}
\providecommand{\BIBdecl}{\relax}
\BIBdecl

\bibitem{LTEASurvey}
I.~F. Akyildiz, D.~M. Gutierrez-Estevez, R.~Balakrishnan, and
  E.~Chavarria-Reyes, ``{LTE-Advanced and the Evolution to Beyond 4G (B4G)
  Systems},'' \emph{Physical Communication (Elsevier) Journal}, vol.~10, pp.
  31--60, {Mar.} 2014.

\bibitem{5G_s2}
{J. G. Andrews, S. Buzzi, W. Choi, S. Hanly, A. Lozano, A. C. Soong, and J. C.
  Zhang}, ``{What Will 5G Be?}'' \emph{IEEE Journal on Selected Areas in
  Communications}, Sept. to appear, 2014.

\bibitem{MassiveMIMO1}
F.~Rusek, D.~Persson, B.~K. Lau, E.~Larsson, T.~Marzetta, O.~Edfors, and
  F.~Tufvesson, ``{Scaling Up MIMO: Opportunities and Challenges with Very
  Large Arrays},'' \emph{IEEE Signal Processing Magazine}, vol.~30, no.~1, pp.
  40--60, Jan. 2013.

\bibitem{PC1}
J.~Jose, A.~Ashikhmin, T.~Marzetta, and S.~Vishwanath, ``{Pilot Contamination
  and Precoding in Multi-Cell TDD Systems},'' \emph{IEEE Transactions on
  Wireless Communications}, vol.~10, no.~8, pp. 2640--2651, Aug. 2011.

\bibitem{PC2}
H.~Huh, S.-H. Moon, Y.-T. Kim, I.~Lee, and G.~Caire, ``{Multi-Cell MIMO
  Downlink With Cell Cooperation and Fair Scheduling: A Large-System Limit
  Analysis},'' \emph{IEEE Transactions on Information Theory}, vol.~57, no.~12,
  pp. 7771--7786, Dec. 2011.

\bibitem{PC7}
F.~Fernandes, A.~Ashikhmin, and T.~Marzetta, ``{Inter-Cell Interference in
  Noncooperative TDD Large Scale Antenna Systems},'' \emph{IEEE Journal on
  Selected Areas in Communications}, vol.~31, no.~2, pp. 192--201, Feb. 2013.

\bibitem{hoydis2013making}
J.~Hoydis, K.~Hosseini, S.~ten Brink, and M.~Debbah, ``{Making Smart Use of
  Excess Antennas: Massive MIMO, Small Cells, and TDD},'' \emph{Bell Labs
  Technical Journal}, vol.~18, no.~2, pp. 5--21, 2013.

\bibitem{FlexTDD}
Z.~Shen, A.~Khoryaev, E.~Eriksson, and X.~Pan, ``Dynamic uplink-downlink
  configuration and interference management in td-lte,'' \emph{Communications
  Magazine, IEEE}, vol.~50, no.~11, pp. 51--59, November 2012.

\bibitem{NomorFlexTDD}
{Nomor Research}, ``{Dynamic TDD for LTE-A and 5G},'' Tech. Rep., Sept. 2015.

\end{thebibliography}

\end{document}